\documentclass[amssymb, amsmath, groupedaddress, reprint]{revtex4-1}

\usepackage{graphicx}
\usepackage[colorlinks=true,linkcolor=blue]{hyperref}
\usepackage{soul,xcolor}
\setstcolor{red}

\begin{document}

\title{Remote detection of rotating machinery with a portable atomic magnetometer}

\author{Luca Marmugi}
\email{l.marmugi@ucl.ac.uk}
\author{Lorenzo Gori}
\author{Sarah Hussain}
\author{Cameron Deans}
\author{Ferruccio Renzoni}

\affiliation{Department of Physics and Astronomy, University College London, Gower Street, London WC1E 6BT, United Kingdom}

\date{Compiled \today}


\begin{abstract}
We demonstrate remote detection of rotating machinery, using an atomic magnetometer at room temperature and in an unshielded environment. The system relies on the coupling of the AC magnetic signature of the target with the spin-polarized, precessing atomic vapor of a radio-frequency optical atomic magnetometer. The AC magnetic signatures of rotating equipment or electric motors appear as sidebands in the power spectrum of the atomic sensor, which can be tuned to avoid noisy bands that would otherwise hamper detection. A portable apparatus is implemented and experimentally tested. Proof-of-concept investigations are performed with test targets mimicking possible applications, and the operational conditions for optimum detection are determined. Our instrument provides comparable or better performance than a commercial fluxgate and allows detection of rotating machinery behind a wall. These results demonstrate the potential for ultrasensitive devices for remote industrial and usage monitoring, security and surveillance.
\end{abstract}


\maketitle


\onecolumngrid
\vskip 5pt
This is a preprint version of the article appeared in Appl.~Opt.~\textbf{56}, 3, 743-749  (2017) DOI: \href{https://doi.org/10.1364/AO.56.000743}{10.1364/AO.56.000743}. \\

One print or electronic copy may be made for personal use only. Systematic reproduction and distribution, duplication of any material in this paper for a fee or for commercial purposes, or modifications of the content of this paper are prohibited.
\vskip 10pt

\twocolumngrid
\section{Introduction}
Remote detection of illicit or hazardous activity inside buildings has become increasingly important in security and defense. At the same time, the possibility of continuous monitoring of machinery would be valuable for industrial activity and quality control. Therefore, the capability of noninvasively and remotely acquiring information about activities in concealed, harsh or inaccessible environments would be an important asset in various fields. Given the broad spectrum of potential applications, including industry, health, and usage monitoring systems (HUMS), defense and security, earth science and aerospace, many different remote sensing approaches, operating either in a passive or active way, have been proposed or developed \cite{wall1, wall2}. Nonetheless, the large civil and industrial use of air conditioning units, turbines, engines and electric motors makes remote detection of rotating machinery and its condition monitoring a very important subject still to be fully addressed \cite{hums01, hums02}. 

In this paper, we demonstrate remote sensing with optical atomic magnetometers (AMs) \cite{budkerbook}, also referred to as optically pumped atomic magnetometers (OPMs). AMs are to date the state-of-the-art for DC and AC magnetic field measurements, even in unshielded environments \cite{bevilacqua2016} and at room temperature \cite{witold}. 

Here, we use a radio-frequency (RF) AM \cite{rfoam1,rfoam2,rfoam3,weis1,weis2} for detecting the rotating target's magnetic signature. In particular, we demonstrate the feasibility of remote detection of rotating machinery, rotating steel samples, and DC and AC electric motors, without any background noise screening. Through-wall detection is also demonstrated. Finally, comparison with a commercial fluxgate highlights the potential of the technique and the proof-of-concept demonstrator we describe in this paper. 

Thanks to the RF AMs' high degree of scalability, potential for miniaturization \cite{chip2004} and wide range of frequency tunability \cite{ghz}, the device we propose is a valid and concrete solution for remote detection of rotating machinery for health, and usage and surveillance applications.   

\begin{figure*}[htbp]
\centering
\includegraphics[width=0.95\linewidth] {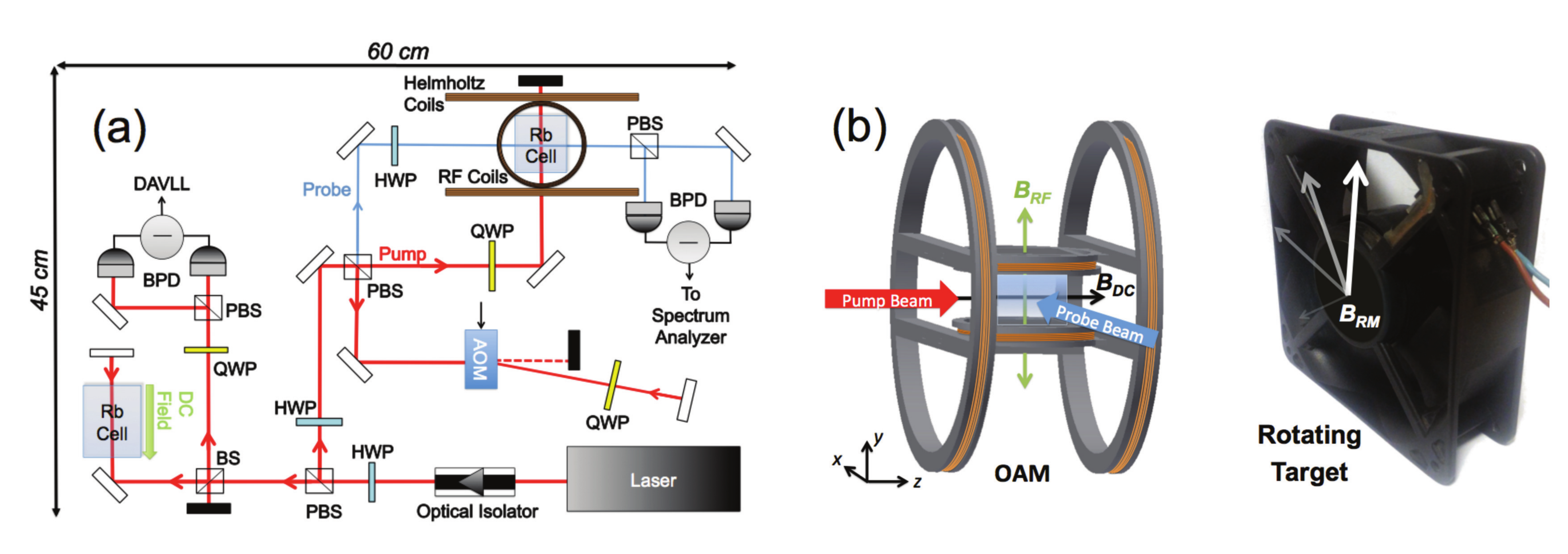}
\caption{Portable RF AM (or OPM) for remote detection of a rotating machinery. \textbf{(a)} Optical setup: a polarizing beam splitter (PBS) splits the laser between the DAVLL and the core of the AM. A second PBS separates the light for the resonant and circularly polarized pump beam and the linearly polarized probe beam, blue-detuned by means of a double-pass acousto-optic modulator (AOM, $\delta_{probe}$= +190 MHz). $\lambda$/2 (HWP) and $\lambda$/4 (QWP) waveplates allow control of the beams' polarizations.  \textbf{(b)} 3D model of the sensor and principle of the detection technique. Atoms are spin-polarized by the pump beam along a DC bias magnetic field ($\mathbf{B_{DC}}$), produced by 136~mm diameter Helmholtz coils with 30 turns. $\mathbf{B_{RF}}$, provided by 54~mm diameter Helmholtz coils with 30 turns, coherently drives atomic precession, probed by the linearly polarized probe beam. The AC field generated by rotating objects or electric motors ($\mathbf{B_{RM}}$) interacts with the precessing atoms, thus modifying the polarization oscillation of the probe beam. The cell's and coils' support is 3D printed in polylactide (PLA). Note that $\mathbf{B_{RM}}$ is indicated here only for illustrative purposes. Its actual orientation depends on the specific target.}
\label{fig:setup}
\end{figure*}

\section{Experimental Setup}
The setup is a table-top (0.40~m $\times$ 0.60~m) portable device, suitable for outdoor operation, based on an optically pumped RF AM [Fig.~\ref{fig:setup}(a)]. 

The core of the sensor is a natural isotopic mixture of Rb atoms contained in a cubic 25 mm glass cell at room temperature. 20 Torr of N$_2$ are added as buffer gas, to increase the light-atom interaction time and limit unwanted depolarization of the atoms.

A RadiantDyes NarrowDiode laser is locked to the $D_2$ line $F=3 \to F'=4$ hyperfine transition of $^{85}$Rb by means of a Doppler-free dichroic atomic vapor laser lock (DAVLL). To reduce costs and footprint, a single laser is used to obtain both the pump beam (for the preparation of the quantum state sensitive to magnetic fields) and the probe beam (for interrogation of the atomic spin precession induced by such fields). In detail, the $\sigma^{+}$ resonant pump beam ($\delta_{pump}=$ 0) aligns the atomic spins along a DC magnetic field, $\mathbf{B_{DC}}=B_{DC}\mathbf{\hat{z}}$, produced by a Helmholtz coil pair [Fig.~\ref{fig:setup}(b)], via optical pumping to the largest angular momentum projection Zeeman sublevel of the ground state, $|F=3, m_F=+3\rangle$. An RF field along $\mathbf{\hat{y}}$, $\mathbf{B_{RF}}$, excites and periodically drives spin-coherences between the ground state Zeeman sublevels. The linear polarization plane of the off-resonant probe beam ($\delta_{probe}=$ +190 MHz) is modulated by the Larmor precession of the atomic spins. The probe's polarization plane is detected by a balanced photodiode (BPD).

In this configuration, the total magnetic field comprises the external magnetic field generated by the rotating machinery or object to be detected, $\mathbf{B_{RM}}$, and the RF field, $\mathbf{B_{RF}}$.

The output of the RF AM is fed to a spectrum analyzer, Anritsu MS2718B. In absence of any external field, the power spectrum is given by a single-narrow peak centered at the frequency of the driving RF field, $f_{RF}$, which can be chosen according to the specific operational conditions and measurement requirements.

\section{Detection Technique}
The technique for the detection of  rotating machinery is shown schematically in Fig.~\ref{fig:setup}(b).  The field $\mathbf{B_{RM}}$, evolving at the frequency $f_{RM}$, perturbs the precession of the atoms. As a result, a mixing is produced within the precessing atomic sample, which couples $\mathbf{B_{RM}}$ and $\mathbf{B_{RF}}$. In the time domain, the resulting AC field is modulated by $B_{RM}$. In the frequency domain, $\mathbf{B_{RM}}$ produces sidebands at $f_{RF}\pm m f_{RM}$, with $m$ integer, in the RF AM's output spectrum. In this way, heterodyne mixing imprints information about the rotating target in extra spectral components of the probe beam's polarization plane oscillation, as a consequence of the non-linear atomic response. The latter allows mixing of oscillating fields rather than the bare sum of their amplitudes. Consequently, the atomic medium exhibits non-trivial response to AC fields, thus altering their spectral components (see, for example, Ref.~\cite{bevilacqua2012}). 

Effectively, the perturbation of the forced trajectories of atomic spins imposed by $\mathbf{B_{RM}}$ can be seen as nutation, which in turn modulates the probe beam's polarization plane rotation. As a result, $f_{RF}$ acts as a carrier for the information stored in the sidebands $f_{RF}\pm m f_{RM}$, thus allowing the decoupling of the RF AM's drive and the information storage and retrieval.

$\mathbf{B_{DC}}$ is used to tune $f_{RF}$, so as to match the operational requirements. In this way, one can avoid noisy bands to increase the detection capability. This feature is of major importance in the case of low-frequency signatures ($<$50 Hz), where $1/f$ noise and poor sensitivity of conventional magnetometers would prevent effective detection. This provides a robust and sensitive detection method for rotating objects and machinery in unshielded and open environments. 

Finally, we note that, even when the object is not rotating ($|\mathbf{B_{RM}}|=0$), our instrument can provide information on conductive targets via the active inductive coupling of magnetic induction tomography \cite{cameron2016, budker2016}. The presence of a conductive object can be inferred from a variation of the amplitude or phase-lag of the detected field.

\section{Experimental Results} 
In this section, results are arranged on the basis of the target under evaluation, or of the specific operational parameters under investigation. We also report on the direct comparison with a commercial fluxgate.

\subsection{Detection of Rotating Objects}
First, we show the detection of rotating metallic objects, with residual magnetization. This is a challenging task for conventional sensors, given the small magnetic signature and low rotational frequency, which would require the sensors to operate in region where the $1/f$ noise is dominant.

A steel disk (diameter 75 mm, height 15 mm) is secured at the center of a spinning gear. The system, realized entirely in plastic to avoid spurious magnetic signatures or eddy current induction, is hand-operated and allows frequencies of the order of $f_{RM}\approx$10 Hz.

Every test is performed at room temperature, in an unshielded environment and without any compensation of the local stray DC magnetic fields, so as to mimic possible field conditions. The AM's frequency is arbitrarily set to $f_{RF}$=10$^{5}$ Hz. Detection of rotation is possible despite $f_{RM} \ll f_{RF}$, as demonstrated below.

\begin{figure}[htbp]
\centering
\includegraphics[width=1\linewidth] {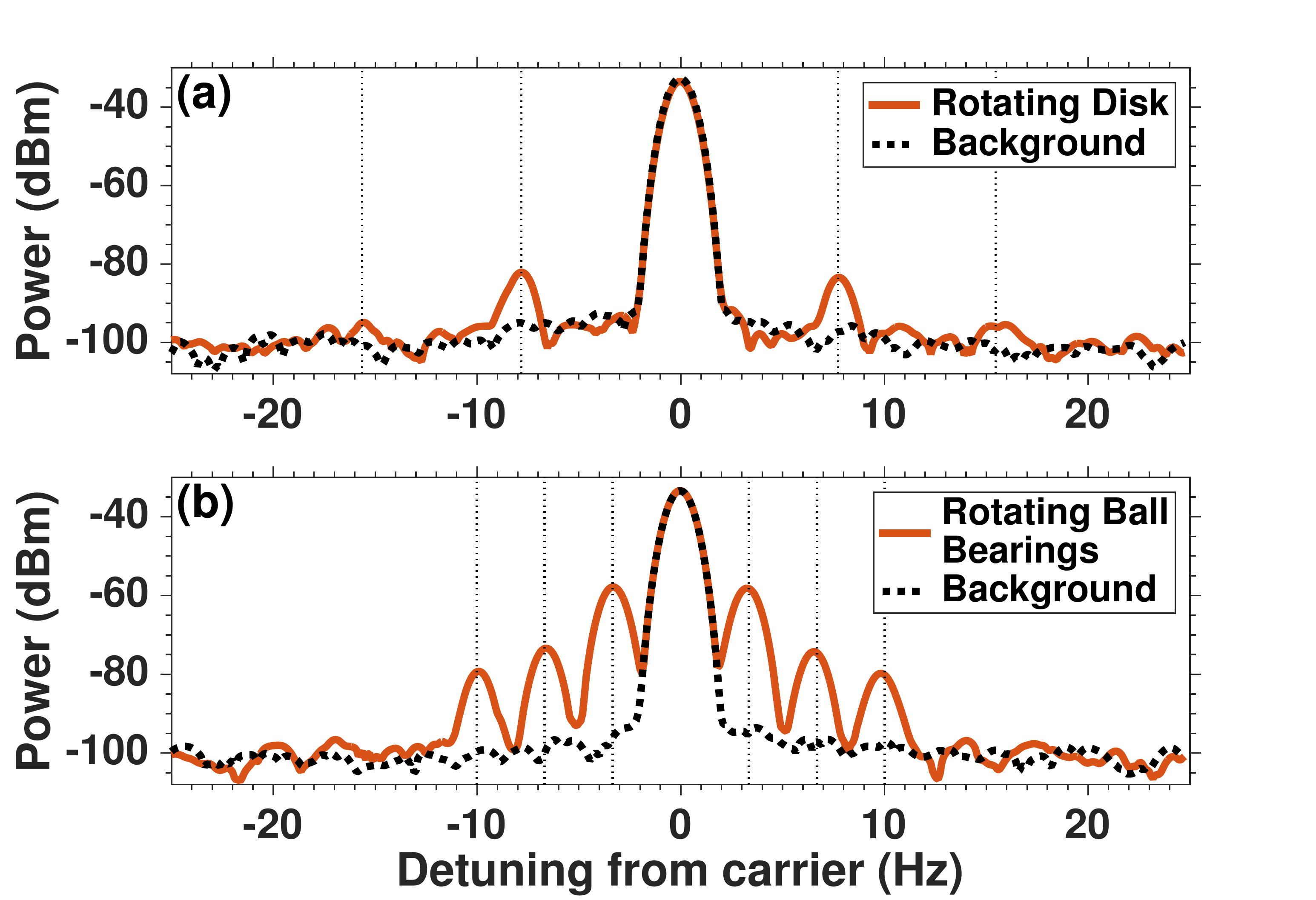}
\caption{Detection of rotating objects. \textbf{(a)} Power density spectrum of the RF AM obtained with a steel disk (75 mm diameter, 15 mm height) rotating around its main axis at $f_{RM}$=(7.7$\pm$0.1) Hz. The disk is placed 0.50 m away from the RF AM. Sidebands and $m$=2 harmonics are marked by the dotted vertical lines. \textbf{(b)} Power density spectrum of the RF AM obtained with ball bearings rotating at $f_{RM}$=(3.3$\pm$0.1) Hz. Sidebands and $m$=2 and $m$=3 harmonics are marked by vertical dotted lines. Target is placed 0.80 m away from the RF AM. Traces are averaged 20 times.}
\label{fig:objects}
\end{figure}

In Fig.~\ref{fig:objects}(a), a typical power spectrum obtained while the steel disk is rotating (continuous line) is compared to the background spectrum, obtained when the object is not rotating (dotted line). 

The large central peak (0 Hz detuning) corresponds to the operational frequency of the RF AM. Sidebands at $\pm f_{RM}$=$\pm$(7.7$\pm$0.1) Hz, about 15 dB larger than the background, reveal the rotation of the object. $m$=2 harmonics are also observed at $\pm$(15.5$\pm$0.1) Hz. The width of the spectral features is increased by small fluctuations in the rotating speed of the disk. At the current stage, the frequency resolution is limited by the bandwidth and the settings of the spectrum analyzer. Therefore, it ranges from 0.1 Hz, as in the case of Fig.~\ref{fig:objects}, to 1 Hz in the broader sweeps.

Results obtained with ball bearings rotating at $f_{RM}$=(3.3$\pm$0.1)~Hz, 0.80 m away from the RF AM, are also shown in Fig.~\ref{fig:objects}(b). In this case, peaks about 35 dB above the background appear at $\pm f_{RM}$. In addition, $m$=2 harmonics at $\pm$(6.7$\pm$0.1) Hz and $m$=3 at $\pm$(10.0$\pm$0.1) Hz are observed well above the background.

\subsubsection{Steel Samples}
Here, we briefly present results concerning small samples of commercially available steels, typically used in manufacturing. We tested three squares of 25 mm $\times$ 25 mm $\times$ 1.5 mm of AISI420, EN19T, and EN24T. Samples are considerably smaller than the previous ones and their homogeneous size and shape allows for direct comparison of the system's performance. Comparable detection performance is observed in all cases.

\begin{figure}[htbp]
\centering
\includegraphics[width=1\linewidth] {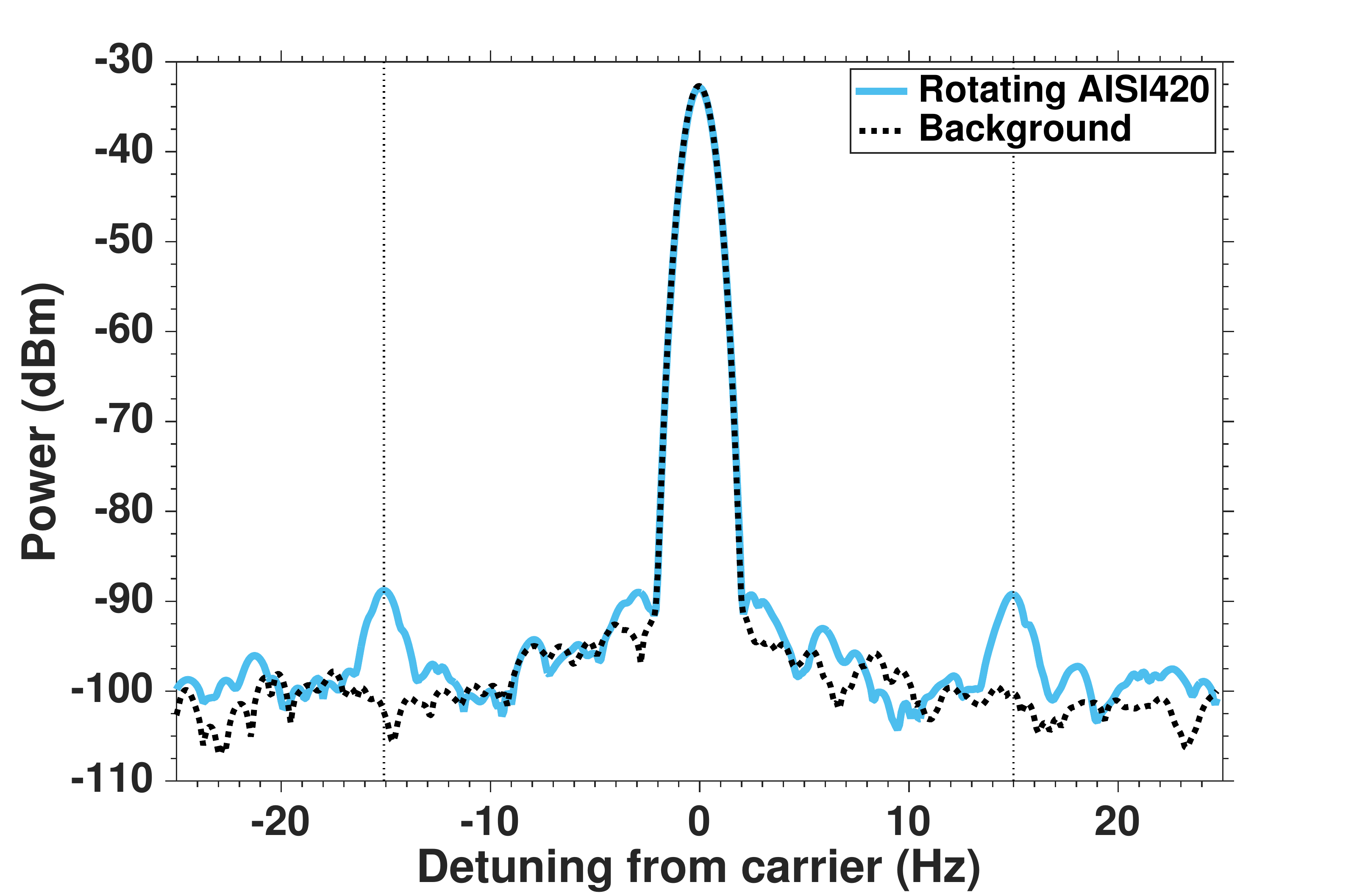}
\caption{Detection of rotating steel: 25 mm $\times$ 25 mm $\times$ 1.5 mm square of AISI420 steel, rotating at 0.40 m from the sensor. The dotted trace marks the background, obtained without the rotating sample. Vertical dotted lines mark the sidebands produced by the rotation at $f_{RM}$=$\pm$(15.1$\pm$0.1) Hz. Traces are the result of 20 averages.}
\label{fig:steels}
\end{figure}

\begin{figure*}[htbp]
\centering
\includegraphics[width=0.62\linewidth]{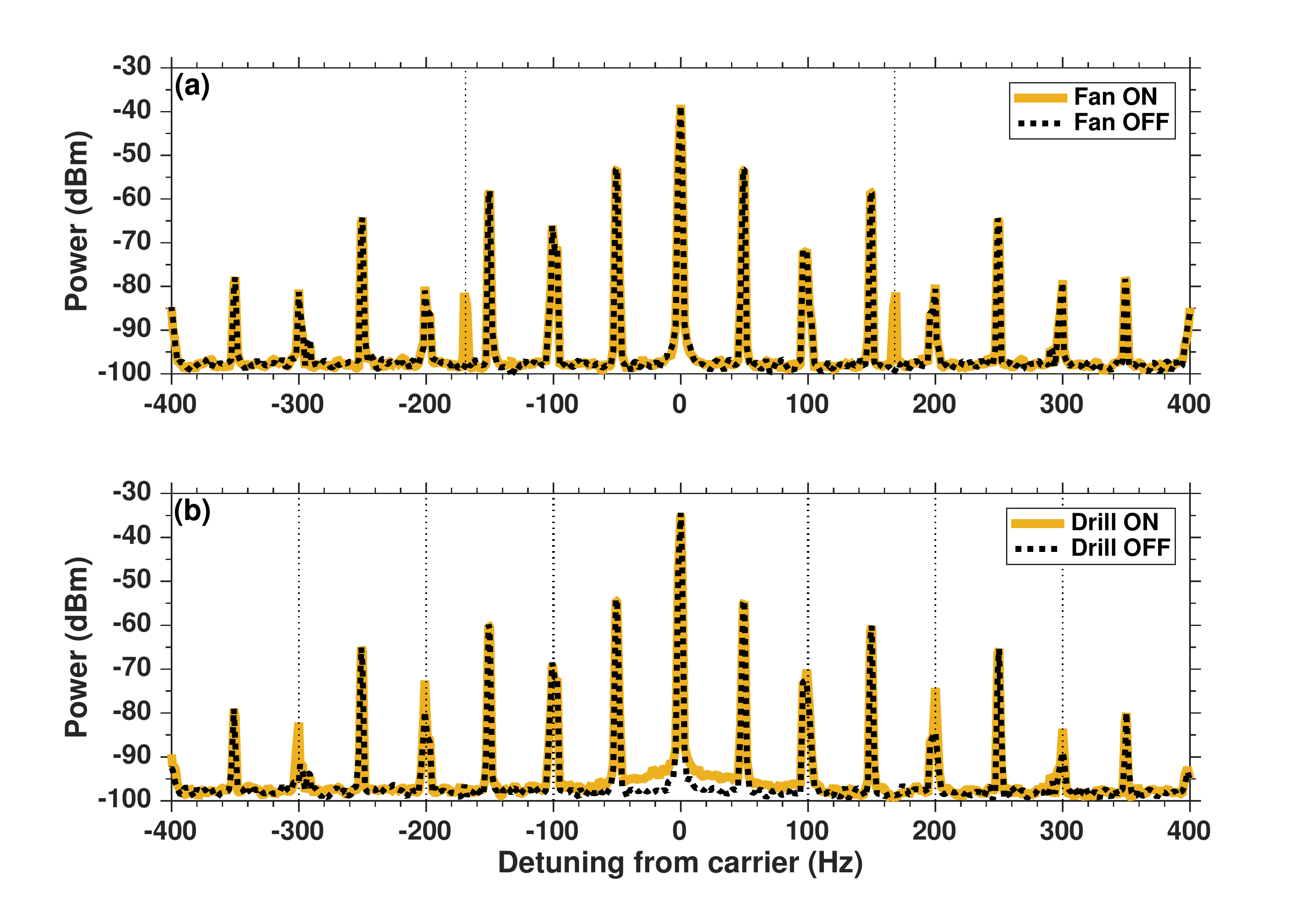}
\caption{Detection of rotating machinery powered by electric motors. \textbf{(a)} Power spectrum of the RF AM obtained with a 24 V DC fan at 1 m from the sensor. \textbf{(b)} Power spectrum of the RF AM obtained with an AC drill at 0.60 m from the sensor. Continuous traces are taken when the motors are on, dotted traces when the motors are off. Vertical lines mark the magnetic signatures of the motors, at $f_{RM}$=$\pm$(170$\pm$1) Hz for the DC case and $f_{RM}$=$\pm$(100$\pm$1) Hz plus $m$=2, 3 harmonics for the AC case. For consistency, all traces are averaged 20 times.}
\label{fig:spectra}
\end{figure*}

Sidebands indicating the rotation of the samples appear in the spectra of all the tested samples, symmetric with respect to the central peak at $f_{RF}$=10$^{5}$ Hz. For the sake of brevity, we report in Fig.~\ref{fig:steels} the results concerning the AISI420 only. In all three cases, the peaks are detected about 10 dB above the background spectrum level, measured when the samples are not rotating.

\subsection{Detection of Electric Motors}
Another important source of magnetic signatures are electric motors \cite{faults}. Here, we test our AM system with both DC and AC motors. 

As a test target mimicking rotating machinery powered by DC motors, a 24 V DC fan (Sunon, model number KDE2412PMB1-6AB) is used. Figure~\ref{fig:spectra}(a) shows spectra obtained with the fan placed 1 m away from the RF AM, after performing an average of 20 measurements.

When the fan is off, the spectrum in Fig.~\ref{fig:spectra}(a) exhibits a set of sidebands either side of the  peak of the carrier at $f_{RF}$=10$^{5}$ Hz. These sidebands, spaced by 50 Hz, are produced by the coupling between the powerline magnetic noise and the RF field.  

When the fan is switched on, two additional sidebands appear at $f_{RM}$=$\pm$(170$\pm$1) Hz. These peaks are the magnetic fingerprint of this particular electric motor. The spectral signature of the DC fan was independently verified with a commercial fluxgate (see also Fig.~ \ref{fig:fg}). It is noteworthy that  this frequency is three times larger than the rotation frequency of the fan, 56$\pm$1 Hz. On the one hand, this confirms that the observed signal is produced by the coupling of $\mathbf{B_{RF}}$ to the AC magnetic field produced by the spinning motor. On the other hand, it provides information about the internal structure of the motor and, potentially, its operational conditions and integrity \cite{faults}. This feature could be of major importance, for example, in the case of continuous monitoring industrial plants and HUMS.
 
As an example of an AC motor, we present in Figure~\ref{fig:spectra}(b) results obtained with a 120 V AC drill (Dumore, model number 37-021) with adjustable speed  (10$^{3}$-1.6$\times$10$^{4}$ RPM). Motor activity is revealed by the increased sidebands at 100 Hz and their $m$=2 and $m$=3 harmonics when the drill is in operation. Despite the overlap with the mainlines peaks, a clear difference is seen with the drill-off spectrum (see in particular the 200 Hz and 300 Hz sidebands). The peaks' frequency is independent of the rotational speed of the drill. A pedestal appears at the carrier's peak, due to low frequency noise produced by the rotating drill head.

\subsection{Response versus RF Power and Target Distance}
Our proof-of-concept system demonstrates that a robust detection of the fan is possible at least up to 2~m, where - in the current testing environment - the signal becomes comparable to the background noise, $P_{BG} \approx -98$~dBm. It is worth underlining that motors with a small size and small magnetic signatures were chosen for the purpose of testing, whereas potential targets are likely to exhibit much larger magnetic signatures.

\begin{figure}[htb]
\centering
\includegraphics[width=1\columnwidth] {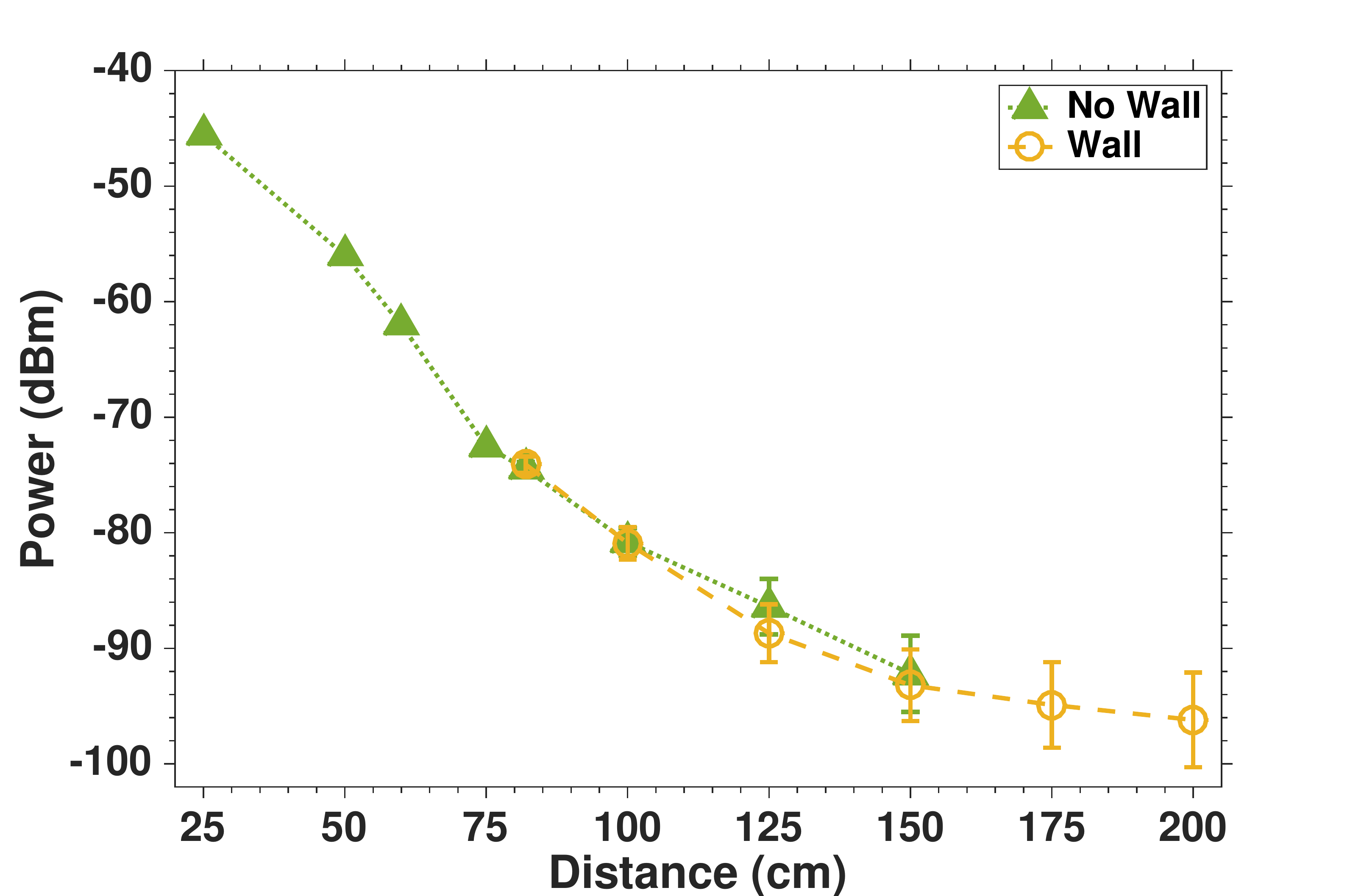}
\caption{DC fan sidebands power as a function of distance from the RF AM, in free space (triangles) and beyond a 0.17~m thick wall (open circles). The uncertainties are the standard deviation of 20 measurements. Lines are guides to the eye. $f_{RF}$=10$^{5}$~Hz; $f_{RM}$=170~Hz.}
\label{fig:vsdistance}
\end{figure}

We have also tested the feasibility of through-wall detection. The rotating motors are placed beyond a 0.17~m thick wall containing concrete, pipes, mainlines, and clutter, 0.70~m away from the RF AM. Results are shown in Fig.~\ref{fig:vsdistance}, which reports the height of the characteristic peak of the fan at frequencies $f_{RF}\pm f_{RM}$ as a function of the distance from the sensor. As demonstrated by the consistency between the two plots in Fig.~\ref{fig:vsdistance}, the wall does not affect the detection.

\begin{figure}[htbp]
\centering
\includegraphics[width=1\linewidth] {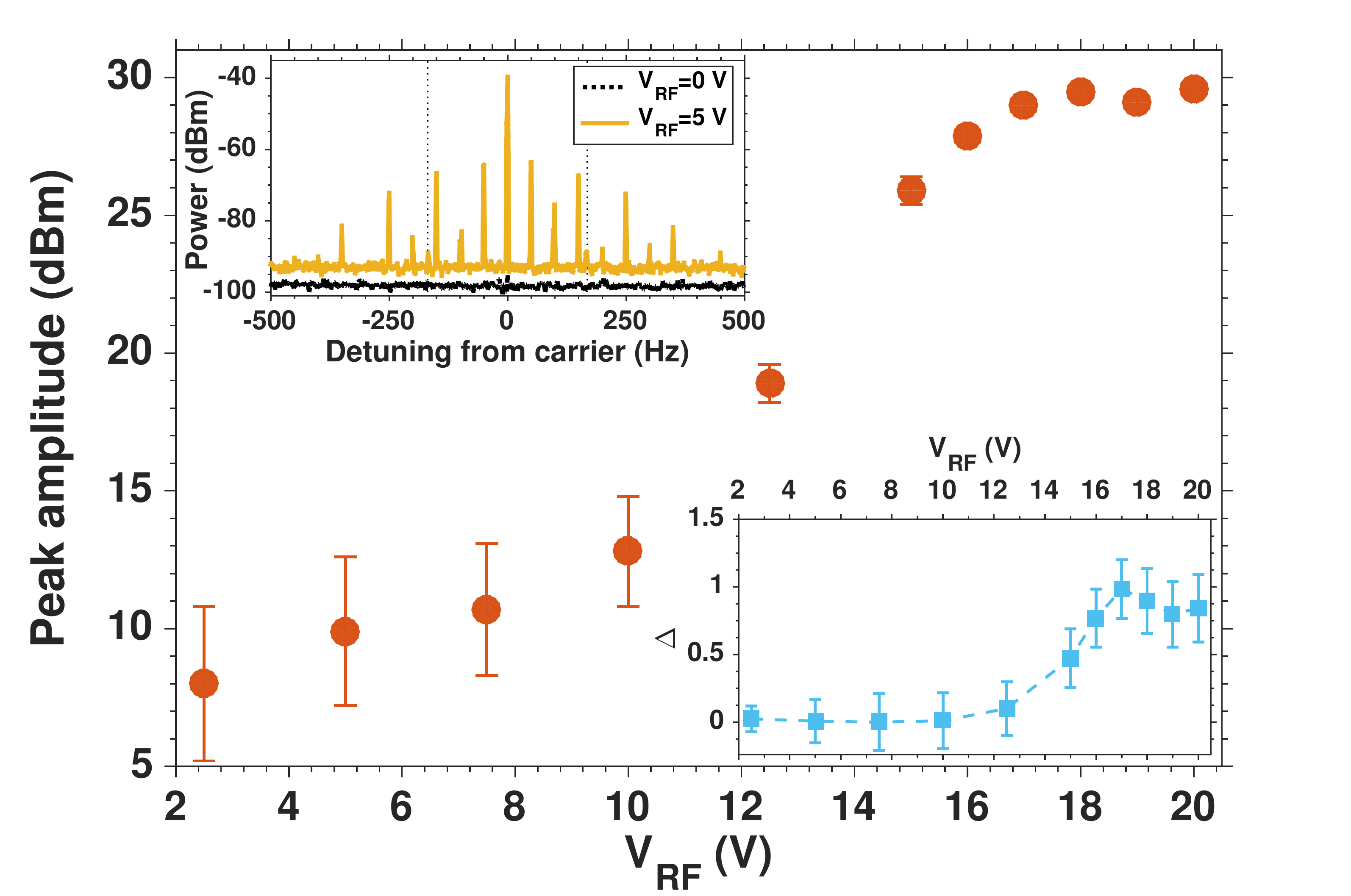}
\caption{DC fan sidebands peak amplitude as a function of $V_{RF}$ supplied to the RF coils at 10$^{5}$ Hz. The fan is 0.75 m from the RF AM. The uncertainties are the standard deviation of 20 measurements.  Top inset: spectra at 0 V (dotted line) and at 5 V (continuous line). For clarity, the spectrum at 5 V has been shifted by +5 dB. The dotted black lines mark the position of the $f_{RM}$= $\pm$(170$\pm$1) Hz sidebands produced by the fan. Bottom inset: relative change $\Delta$ of the carrier's peak power in the corresponding cases. The dashed line is a guide for the eye.}
\label{fig:vspower}
\end{figure}

In Fig.~\ref{fig:vspower} we show the effect of the driving amplitude of the RF field ($V_{RF}$) at 10$^{5}$~Hz. The spectrum is flat when the RF supply is off ($V_{RF}$=0), while the central peak and sidebands appear as soon as the voltage increases (see top inset). Given the proof-of-principle nature of the present work and taking into consideration data reported in literature on atomic magnetometers \cite{rfoam1,rfoam2,rfoam3,weis1,weis2}, one can anticipate a great potential for a further increase in detection capability and range.

When $V_{RF}\neq$0, the height of the fan peak progressively increases and saturates at about $V_{RF}$=17~V, chosen as the RF amplitude for all the measurements presented here, unless otherwise stated. With this $V_{RF}$, the pure optical rotation signal as produced by the AM is about 30~dB at $f_{RF}$, where direct coupling of the RF to the photodiode's electronics produces a spurious contribution of 30 dB higher than the background level.

Thanks to the detection method, this is not a problem: as a consequence of the decoupling of the information storage in the sidebands from the sensor's drive, extra contributions at $f_{RF}$ will not have a direct detrimental effect on the signals of interest.

In addition, Fig.~\ref{fig:vspower} demonstrates the role of $\mathbf{B_{RF}}$ for the detection of $\mathbf{B_{RM}}$. It thus further confirms the analysis in terms of fields' mixing inside the atomic sample. In detail, given the identical conditions of the measurements and the constant average amplitude of $|\mathbf{B_{RM}}|$, the increase of the $f_{RF}\pm f_{RM}$ peaks' amplitude as a function of $V_{RF}$ (Fig.~\ref{fig:vspower}) can be explained only by a direct dependence of the sidebands on the carrier's amplitude. This is a well-known feature of heterodyne mixing, whereby a secondary frequency component (in this case $f_{RM}$) is effectively amplified in a sideband ($f_{RF}\pm f_{RM}$) when the power of the carrier frequency ($f_{RM}$) is increased.

As supporting evidence, we show in the bottom inset of Fig.~\ref{fig:vspower} the relative change $\Delta_{i}=(P_{i}-P_{min})/P_{max}$ of the carrier's peak at $f_{RF}$ as a function of $V_{RF}$ in the same experiment. Here, $P_{i}$ is the power of the peak at $f_{RF}$, and $P_{min}$ and $P_{max}$ are the minimum and the maximum $P_{i}$, respectively. $\Delta_{i}$ behavior  is consistent with the observed increase of the sidebands' amplitude. Its relative change  is consistent with the observed increase of the sidebands' amplitude. In other words, as the carrier peak increases, the sidebands' amplitude increases, enhancing the detection capability and extending its range.

In summary, to maximize the amplitude of the sidebands and hence of the likelihood of detection, one should over-drive the RF AM. In the present case, this implies $V_{RF}>$ 16~V. In usual conditions, however, this would effectively reduce the sensitivity at the carrier frequency \cite{smullin2009}. Nevertheless, because of the approach demonstrated here, this detrimental effect has no impact on the detection of the rotating machinery's magnetic signatures. In other words, the optimum sidebands detection can be safely pursued, even at the expenses of the absolute sensitivity of the RF AM.

\subsection{Comparison with Commercial Fluxgate}
Finally, to validate our instrument and its operational modality, we present a comparison of the results concerning the DC fan obtained with a commercial fluxgate and with our RF AM.

\begin{figure}[htbp]
\centering
\includegraphics[width=1\linewidth] {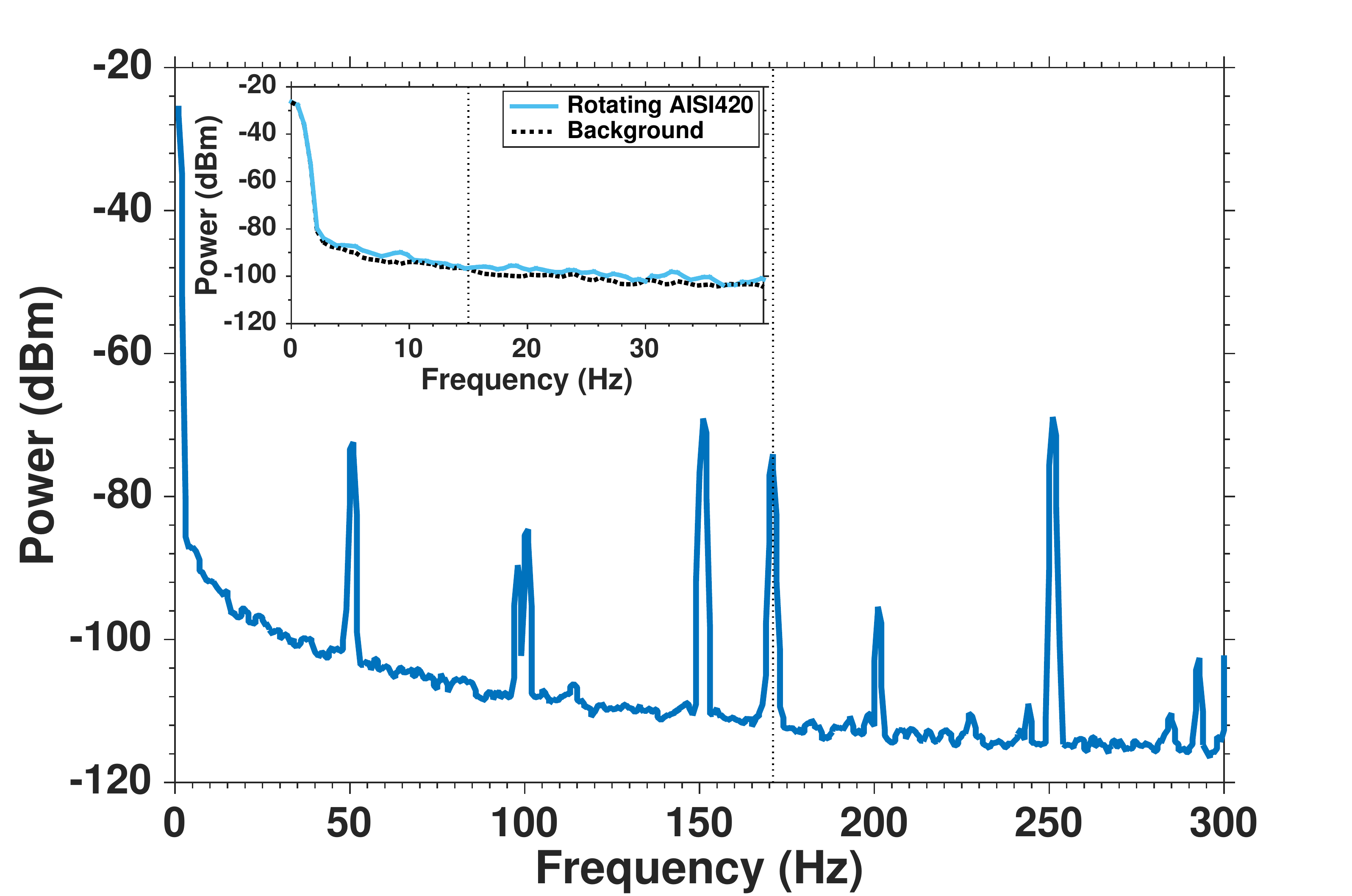}
\caption{DC fan signature monitored with a commercial fluxgate. Vertical dotted line marks the position of the $f_{RM}=($170$\pm$1) Hz component produced by the rotating DC motor, placed 0.50 m away from the sensor. Inset: AISI420 square rotating at $f_{RM}=(15.0\pm0.1)$~Hz, monitored with a commercial fluxgate placed 0.40 m away from the sensor. No signature is found at $f_{RM}$, marked by a dotted vertical line. Traces are averaged 20 times.}
\label{fig:fg}
\end{figure}

In Fig.~\ref{fig:fg} we show the spectrum obtained by connecting the fluxgate's output to the spectrum analyzer, in the same configuration as that used with our RF AM. The DC fan is placed 0.50 m away from the fluxgate. After 20 averages, a single $f_{RM}$=(170$\pm$1)~Hz peak is clearly visible, as well as the 50~Hz noise produced by power lines and its harmonics. This independently confirms that the peaks detected by the RF AM are produced by the DC fan.

It is important to underline that the $f_{RM}$ spectral component is obtained here by \emph{direct} detection. The fluxgate cannot support any frequency mixing, as demonstrated below. The DC fan peak lies on top of an increasing background, which rapidly exceeds -90~dBm below 10~Hz, a consequence of the $1/f$ noise of both the fluxgate and the spectrum analyzer. 

This is detrimental for low-frequency measurements. For example, detection of the sidebands produced by steel samples such as those shown in Fig.~\ref{fig:steels} would be impossible (see inset of Fig.~\ref{fig:fg} in the case of AISI420). First, the lack of mixing with a carrier causes a reduction in the peaks' amplitude. Second, the peaks are comparable with the noise floor.

\begin{figure}[htbp]
\centering
\includegraphics[width=1\linewidth] {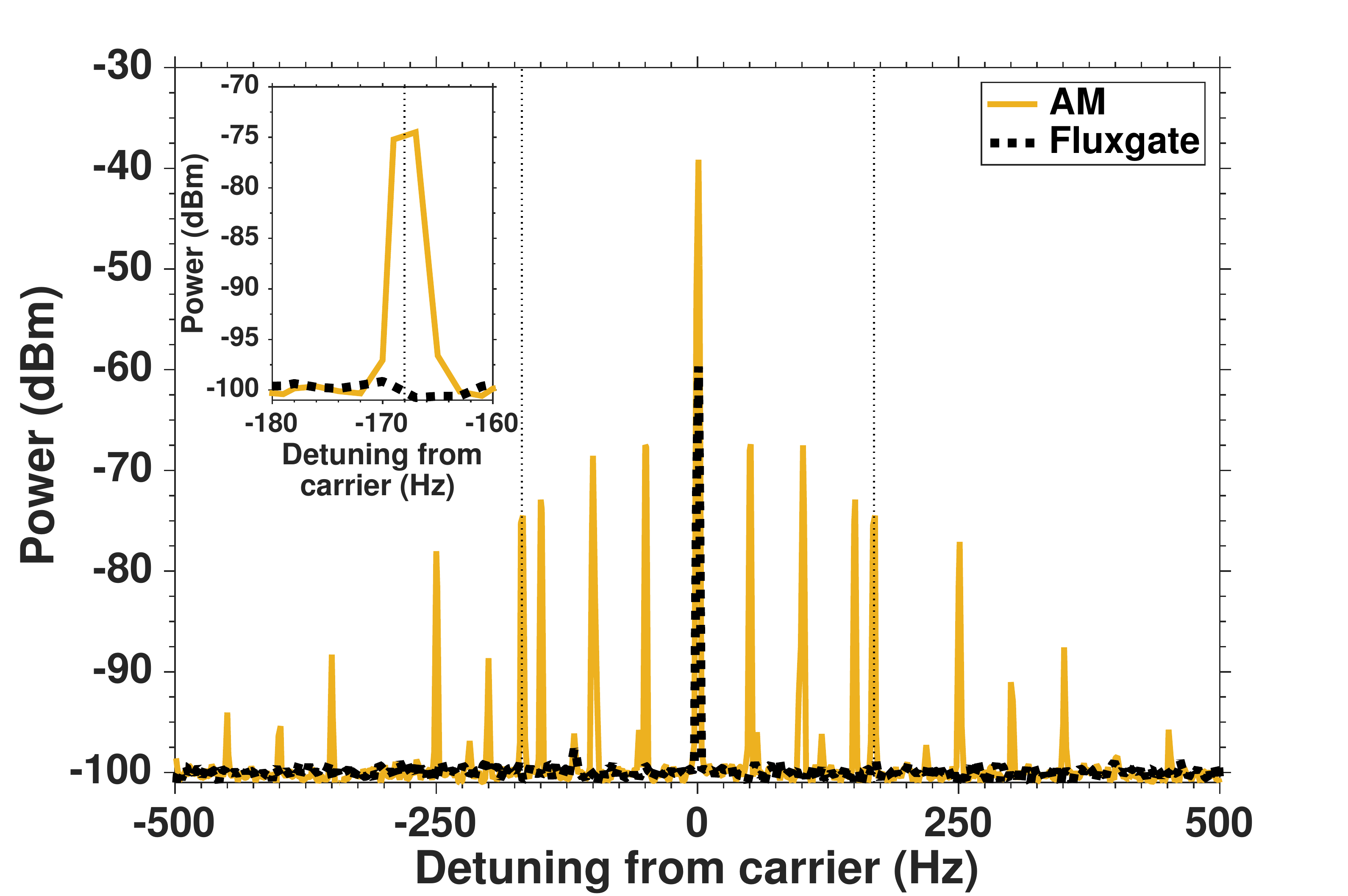}
\caption{Comparison of the power density spectra produced by the 24 V DC fan at $f_{RF}$=3$\times$10$^{4}$ Hz obtained with the RF AM (continuous line) and a commercial fluxgate (dotted line). Sidebands at $f_{RM}$=$\pm$(170$\pm$1) Hz are visible only in the atomic magnetometer trace. In the inset, a detailed view of the $-f_{RM}$=$-$170~Hz peak is shown. An offset is added to both spectra in order to overlap the traces.}
\label{fig:comparison}
\end{figure}

In Fig.~\ref{fig:comparison} the output spectra of both the fluxgate and the RF AM are shown at $f_{RF}=3\times 10^{4}$ Hz. This frequency was chosen in order to ensure operation of both sensors and, therefore, the possibility of direct comparison of their response. For this test, the fluxgate is placed in the same position as the vapor cell of the RF AM.

Upon inspection of Fig.~\ref{fig:comparison}, it is clear how detection of $f_{RF}$ - in the current settings and working conditions - is more effective in the case of the AM.
More importantly, in view of the detection of rotating machinery, no sidebands are detected in the fluxgate's output. This is a consequence of the lack of frequency mixing in the latter and thus a further confirmation of the atoms' role in the creation of sidebands.

On the contrary, well-defined sidebands can be observed in the AM trace, including 35 dBm high peaks at $f_{RM}$=$\pm$(170$\pm$1) Hz. This is consistent with the previous observations, despite the change in $f_{RF}$.

\section{Conclusions}
In conclusion, we have realized a proof-of-concept demonstration of remote detection of rotating objects and of both DC and AC electric motors with a compact and portable optically pumped atomic magnetometer. The magnetic signatures of rotating targets appear as symmetric sidebands of the carrier which drives the RF AM and can be thus detected also in critical conditions, such as low frequency, where conventional sensors usually have poor performance. 

The proposed instrument takes advantage of the performance of optically pumped atomic magnetometers and the non-linear nature of atomic response to oscillating fields. The use of an RF AM, in particular, offers a broad range of tunability, which is of major importance for operation in unscreened scenarios and different environmental noise conditions. In this way, by using a less noisy frequency band and frequency mixing amplification, our proof-of-concept demonstrator exhibits comparable performance to that of commercial devices in the case of high frequency rotation ($\geq$50 Hz), and increasingly better response in the case of low frequency rotation ($<$50 Hz). In particular, our instrument detects rotation of objects and machinery below 15 Hz.

Furthermore, the demonstrated detection of rotating machinery through concrete walls makes this approach suitable for harmless, noninvasive and nondisrupting remote sensing for security and surveillance, but also for continuous control of civil industrial processes and health and usage monitoring. 

The proposed system is therefore a promising candidate for successful deployment in real-life scenarios, thanks to its portability, its margin for further improvement detection capability and range, and its potential for miniaturization and power consumption containment.

\vskip 2pt
The work was supported by DSTL and CDE in the framework of ``What's inside that building?'' program. L.~G. is supported by Wellcome Trust. S.~H. is supported by DSTL -- Defence and Security PhD -- Sensing and Navigation using Quantum 2.0 technology. C.~D. is supported by the Engineering and Physical Sciences Research Council $[$grant number EP/L015242/1$]$. We thank D.~Harvey and D.~Tee (THALES UK Research and Technology) for their support.

\bibliography{AO_UCL}

\end{document}